\documentclass{article}
\usepackage{amsfonts}

\newcommand{\beq}{\begin{equation}}
\newcommand{\eeq}{\end{equation}}

\begin{document}

\title{On generalisations of Calogero-Moser-Sutherland quantum problem and WDVV equations}

\maketitle

\begin{center}

{\bf A.P.Veselov }


\bigskip

{\it Department of Mathematical Sciences, Loughborough University,
Loughborough, Leicestershire, LE 11 3TU, UK
}

\bigskip

{\it Landau Institute for Theoretical Physics, Kosygina 2,

 Moscow, 117940,  Russia

\bigskip

e-mail: A.P.Veselov@lboro.ac.uk,
}

\end{center}

\bigskip

\bigskip

{\small  {\bf Abstract.}
It is proved that if the Schr\"odinger equation $L \psi = \lambda \psi$ of Calogero-Moser-Sutherland type with
$$L = -\Delta + \sum\limits_{\alpha\in{\cal A}_{+}} \frac{m_{\alpha}(m_{\alpha}+1)
(\alpha,\alpha)}{\sin^{2}(\alpha,x)}$$
has
a solution of the product form
$\psi_0 = \prod_{\alpha \in {\cal {A}_+}} \sin^{-m_{\alpha}}(\alpha,x),$ 
then the function
$F(x) =\sum\limits_{\alpha \in \cal {A}_{+}} m_{\alpha} (\alpha,x)^2 \, {\rm log} \, 
(\alpha,x)^2$
satisfies the generalised WDVV equations.}

\bigskip

\section*{Introduction.}

Let $\cal{A}$ be a finite set of vectors $\alpha$ in 
the Euclidean space ${\bf R}^n$ which generates the space and is invariant under the symmetry $x \rightarrow -x.$
We assume that $-\alpha$ is the only vector from $\cal{A}$ which is proportional to $\alpha$.
Let $\cal{A}_{+}$ be its half positive with respect to some linear form.
Let us prescribe to each vector $\alpha \in \cal{A}$ a real number ("multiplicity") $m_{\alpha}$
such that $m_{-\alpha}= m_{\alpha}.$ 

Consider the following Schr\"odinger operator

\begin{equation}
\label{1}
L = -\Delta + \sum\limits_{\alpha\in{\cal A}_{+}} \frac{m_{\alpha}(m_{\alpha}+1)
(\alpha,\alpha)}{\sin^{2}(\alpha,x)}.
\end{equation}
When ${\cal {A}}$ is a root system with the multiplicities invariant under the corresponding Weyl group $W$
this is an integrable generalisation of the Calogero-Moser-Sutherland (CMS) operator \cite{C,S} suggested by 
Olshanetsky and Perelomov  \cite{OP}. As it has been shown in \cite{VFCh, ChFV1}
there are also non-symmetric integrable generalisations of this problem. This discovery led to the notion of the
locus configurations which play a crucial role in the theory of the Huygens' principle \cite{ChFV2}.

It turned out \cite{V1,V2} that the locus configurations
discovered in \cite{VFCh, ChFV1, ChFV2} can be used also to construct new solutions of the generalised 
WDVV (Witten-Dijgraaf-Verlinde-Verlinde) equations:
\beq
F_iF_k^{-1}F_j=F_jF_k^{-1}F_i, \quad i,j,k=1,\ldots,n.
\label{wdvv}
\eeq
Here $F_m$ is the $n\times n$ matrix constructed from the third partial
derivatives of the unknown function $F=F(x^1,\ldots,x^n)$:
\beq
\label{f}
(F_m)_{pq}=\frac{\partial ^3 \, F}{\partial x^m \partial x^p \partial x^q},
\eeq
In this form these equations have been written by Marshakov, Mironov and Morozov, 
who showed that the Seiberg-Witten prepotential in $N=2$ four-dimensional 
supersymmetric gauge theories satisfies this system \cite{MMM}. Originally these equations 
have been introduced first
in topological field theory as some associativity conditions \cite{W,DVV}
and have been later investigated in this context by Dubrovin \cite{D}.

However in \cite{ChV} we have shown that the relation discovered in \cite{V1}
between the locus configurations and WDVV equations does not always work,
which raised the question what exactly is the property behind this relation.

The aim of this paper is to present the answer to this question. 
The main result is the following

{\bf Theorem.} {\it If the Schr\"odinger equation $L \psi = \lambda \psi$
of Calogero-Moser-Sutherland type (\ref{1}) has a solution of the product form
$$\psi_0 = \prod_{\alpha \in {\cal {A}_+}} \sin^{-m_{\alpha}}(\alpha,x),$$ 
then the function
$$F(x) =\sum\limits_{\alpha \in \cal {A}_{+}} m_{\alpha} (\alpha,x)^2 \, {\rm log} \, 
(\alpha,x)^2$$
satisfies the generalised WDVV equations.}

We should mention that the eigenfunctions of the product form play a special role
in the theory of the original CMS problem giving the ground states of the system \cite{C,S}.
For the deformed CMS problem this becomes more complicated because these solutions
become singular on some hyperplanes, so we can consider these solutions as the eigenfunctions 
only in the formal sense.

\section*{Special class of solutions to generalised WDVV equations}

In this section we essentially follow the analysis of the special 
solutions to WDVV equations from \cite{V1,V2}. 

It is known \cite{MMM,MM}  that WDVV
equations (\ref{wdvv}), (\ref{f}) are equivalent to the equations
\beq
F_iG^{-1}F_j=F_jG^{-1}F_i, \quad i,j=1,\ldots ,n,
\label{fgf}
\eeq
where $G=\sum\limits_{k=1}^n\eta^kF_k$ is any particular invertible linear combination of $F_i$
with the coefficients, which may depend on $x$. Introducing the matrices $\check F_i=
G^{-1}F_i$ one can rewrite (\ref{fgf}) as the commutativity relations 
\beq
\left[ \check F_i, \check F_j \right] =0, \quad i,j=1,\ldots ,n,
\label{com}
\eeq

We will consider the following particular class of the solutions to these equations.

Let $V$ be a real vector space of dimension $n$, $V^*$ be its dual space consisting of 
the linear functions on $V$ (covectors),  $\mathfrak{A}$ be a finite
 set of noncollinear covectors $\alpha \in V^*$ generating $V^*$. 

Consider the following function on $V$:
\beq
F^{\mathfrak{A}}=\sum\limits_{\alpha \in \mathfrak{A}} (\alpha,x)^2 \, {\rm log} \, 
(\alpha,x)^2,
\label{mF}
\eeq
where $(\alpha,x)=\alpha(x)$ is the value of covector $\alpha \in V^*$ on a vector $x\in V$. 
For any basis $e_1,\ldots, e_n$ we have the corresponding coordinates
 $x^1,\ldots , x^n$
in $V$ and the matrices $F_i$ defined according to  (\ref{f}). In a more invariant form
 for any vector $a \in V$ one can define the matrix 
$$
F_a=\sum\limits_{i=1}^n a^iF_i.
$$
One can easily check that
 $F_a$  is the matrix 
of the following bilinear form on  $V$
$$
F_a^{\mathfrak{A}}=\sum\limits_{\alpha \in \mathfrak{A}} \frac{(\alpha,a)}{(\alpha,x)}
\alpha\otimes \alpha,
$$
where $\alpha\otimes \beta (u,v)=\alpha (u)\beta (v)$ for any $u,v \in V$ and $\alpha, \beta \in V^*$.
Define $G$ 
as $F_x$, i.e.
$$
G=\sum\limits_{i=1}^n x^iF_i
$$
which is actually the matrix of the bilinear form
\beq
G^{\mathfrak{A}}=\sum\limits_{\alpha \in \mathfrak{A}} \alpha\otimes \alpha,
\label{mG}
\eeq
which does not depend on $x$.
Since the covectors $\alpha \in \mathfrak{A}$ generate $V^*$,
the form $G^{\mathfrak{A}}$ is non-degenerate. 
 This means  that the natural linear mapping
$\varphi_\mathfrak{A} : V\rightarrow V^*$  defined by the formula 
$$
(\varphi_\mathfrak{A}(u),v)=G^{\mathfrak{A}} (u,v), \, u,v \in V
$$
is invertible. We will denote  $\varphi_\mathfrak{A}^{-1}(\alpha),\,  \alpha \in V^*$ as
$\alpha^{\vee}$. By definition 
$$\sum\limits_{\alpha \in \mathfrak{A}} \alpha^{\vee}\otimes \alpha = Id$$
as an operator in $V^*$ or equivalently
\beq
(\alpha,v)=\sum\limits_{\beta \in \mathfrak{A}}(\alpha, \beta^{\vee})(\beta,v).
\label{vee}
\eeq
for any $\alpha \in V^*, v \in V$.
Now according to (\ref{com}) the WDVV equations (\ref{wdvv},\ref{f}) for the function (\ref{mF}) 
can be rewritten as
\beq
\label{ab}
\left[\check F_a^\mathfrak{A}, \check F_b^\mathfrak{A} \right] =0
\eeq
for any $a,b \in V$, where the operators $\check F_a^{\mathfrak{A}}$ are defined as
\beq
\check F_a^{\mathfrak{A}}=\sum\limits_{\alpha \in \mathfrak{A}} \frac{(\alpha,a)}
{(\alpha,x)}\alpha^\vee \otimes \alpha.
\eeq
Let $G_\mathfrak{A}$ be the form on $V^*$ induced by $G^\mathfrak{A}$:
$$G_\mathfrak{A} (\alpha, \beta) = G^\mathfrak{A} (\alpha^\vee, \beta^\vee).$$
Its matrix in the corresponding basis is $G^{-1}.$
A simple calculation  shows that (\ref{ab}) can be rewritten as 
\beq
\label{sc}
\sum\limits_{\alpha \ne \beta, \alpha,\beta \in \mathfrak{A}}
\frac{G_\mathfrak{A} (\alpha, \beta)B_{\alpha,\beta}(a,b)}
{(\alpha,x)(\beta,x)}\alpha\wedge \beta \equiv 0,
\eeq
where 
$$
\alpha\wedge \beta=\alpha\otimes \beta-\beta \otimes \alpha
$$ 
and
$$
B_{\alpha,\beta}(a,b)=\alpha\wedge \beta(a,b)=\alpha(a)\beta(b)-\alpha(b)\beta(a).
$$

One can check that the relations (\ref{sc}) are equivalent to
$$
\sum\limits_{\beta \neq \alpha, \beta \in \Pi \cap \mathfrak{A}}
\frac{G_\mathfrak{A} (\alpha, \beta) B_{\alpha,\beta}(a,b)}
{(\beta,x)}\alpha \wedge \beta |_{(\alpha,x)=0}\equiv 0
$$
for any $\alpha \in \mathfrak{A}$ and any two-dimensional plane $\Pi$ containing $\alpha,$
and therefore to
\beq
\label{pi}
\sum\limits_{\beta \neq \alpha, \beta \in \Pi \cap \mathfrak{A}}
G_\mathfrak{A} (\alpha, \beta) B_{\alpha,\beta}(a,b) = 0
\eeq
for any such $\alpha$ and $\Pi$.

Summarising we have the following result (cf. \cite{V1, V2}).

{\bf Proposition 1.} {\it The function (\ref{mF}) satisfies the generalised WDVV equations
(\ref{wdvv},\ref{f}) iff the relations (\ref{pi}) are satisfied for any $\alpha \in \mathfrak{A}$ 
and any two-dimensional plane $\Pi$ such that $\alpha \in \Pi$.}

{\bf Remark.} In \cite{V1, V2} the conditions (\ref{pi}) have been reformulated in geometric 
terms ($\vee$-conditions). For the purpose of the present paper it is more convinient to use
these conditions in the original form.

Any Coxeter root system satisfies these conditions but there are many more examples
(see \cite{V1,V2, ChV}).

\smallskip

\section*{Main Identity and proof of the Theorem.}

Let now $\cal{A}$ be a finite set of the vectors $\alpha$ in 
the Euclidean space ${\bf R}^n$ with multiplicities $m_{\alpha}$.
We will assume that this set is invariant under the symmetry $x \rightarrow -x$
and that $-\alpha$ is the only vector from $\cal{A}$ which is proportional to $\alpha$.
By ${\cal A}_{+}$ we mean the positive half of $\cal{A}$:
$${\cal A}_{+} = \lbrace \alpha \in {\cal A} : (\alpha, v) > 0 \rbrace$$
for some $v \in {\bf R}^n$ which is generic in the sense that $(\alpha, v) \neq 0$ for all $\alpha \in {\cal A}$.

Consider the Schr\"odinger equation

\begin{equation}
\label{Schr}
L \psi = \lambda \psi,
\end{equation}
where
$$L = -\Delta + \sum\limits_{\alpha\in{\cal A}_{+}} \frac{m_{\alpha}(m_{\alpha}+1)
(\alpha,\alpha)}{\sin^{2}(\alpha,x)}.$$
Let us look for the solutions of this equation of the form
\begin{equation}
\label{fun}
\psi_0 = \prod_{\alpha \in {\cal {A}_+}} \sin^{-m_{\alpha}}(\alpha,x).
\end{equation}
By straightforward calculation one can check the following result (cf. \cite{OP2}).

{\bf Lemma 1.} {\it The function (\ref{fun}) satisfies the equation (\ref{Schr})
for some $\lambda$ iff the following Main Identity holds:}
\begin{equation}
\label{Main}
\sum\limits_{\alpha \ne \beta, \alpha,\beta\in {\cal A}_{+}} m_{\alpha}m_{\beta}(\alpha, \beta) (\cot(\alpha,x) \cot(\beta,x)+1) \equiv 0.
\end{equation}
{\it The eigenvalue $\lambda$ in that case has a form}
$\lambda = |\rho(m)|^2$ {\it where}
$\rho(m) = \sum\limits_{\alpha \in {\cal {A}_+}} m_{\alpha} \alpha.$

Actually first we have the condition that the sum
$$S = \sum\limits_{\alpha \ne \beta, \alpha,\beta\in {\cal A}_{+}} m_{\alpha}m_{\beta}(\alpha, \beta) \cot(\alpha,x) \cot(\beta,x)$$
must be a constant in $x$. Considering then $x=i t v$ with a large positive $t$ and $v$ from the definition of ${\cal A}_{+}$
we see that this constant must be equal to 
$$S = -\sum\limits_{\alpha \ne \beta, \alpha,\beta\in {\cal A}_{+}} m_{\alpha}m_{\beta}(\alpha, \beta).$$
This leads also to the formula for the eigenvalue $\lambda.$ It is worthy to note that this formula 
implies that the vector $\rho(m)$ which depends on the choice of the positive part ${\cal A}_{+}$ 
(i.e. on the choice of  $v$) has the norm $|\rho(m)|$ independent on this choice.

{\bf Remark.} The gauged operator $\tilde {L} = \hat \psi_0^{-1} L \hat \psi_0$, where $\hat \psi_0$ is the operator of 
multiplication by $\psi_0$, in this case has the form $\tilde{L} = - L^{rad} + |\rho(m)|^2,$
$$L^{rad} = \Delta - 2 \sum\limits_{\alpha\in{\cal A}_{+}} m_{\alpha} \cot (\alpha,x) \partial_{\alpha}.$$
When $\cal{A}$ is a root system with special multiplicities $L^{rad}$ is the radial part of Laplace-Beltrami operator
on some symmetric space \cite{Hel}. Thus, the property of $L$ we discuss is equivalent to the existence
of the "radial gauge". We should mention that the relation between the radial parts of the Laplace operators
and quantum many-body problems was first observed (in the case of the symmetric spaces $SU(n)/SO(n)$)
by Berezin, Pokhil and Finkelberg \cite{BPF} and was investigated in details 
by Olshanetsky and Perelomov in \cite{OP2}. 

The Main Identity is equivalent to the following identities:
\begin{equation}
\label{k}
\sum\limits_{\beta\in{\cal A}_{+}, \beta \ne \alpha} m_{\alpha}m_{\beta} (\alpha, \beta) \cot (\beta,x)|_{(\alpha,x) = k \pi} \equiv 0
\end{equation}
for each $\alpha \in {\cal A}_+$ and $k \in {\bf Z}$.
In particular,

\begin{equation}
\label{0}
\sum\limits_{\beta\in{\cal A}_{+}, \beta \ne \alpha} m_{\alpha}m_{\beta} (\alpha, \beta) \cot (\beta,x)|_{(\alpha,x)=0} \equiv 0.
\end{equation}
must be satisfied on all the hyperplanes $(\alpha, x) =0.$
It is easy to see that the last identities are equivalent to the set of identities
\begin{equation}
\label{plane}
\sum\limits_{\beta\in{\cal A}_{+}\cap \Pi, \beta \ne \alpha} m_{\alpha}m_{\beta} (\alpha, \beta) \cot (\beta,x)|_{(\alpha,x)=0}\equiv 0
\end{equation}
hold for any plane $\Pi$ containing vector $\alpha \in {\cal A}_{+}.$

Let's look now at these two-dimensional identities. Let $\Pi$ be the plane generated by
$\alpha,\beta \in {\cal A}_{+}, \alpha \ne \beta$. We can assume that $(\alpha, \beta) \ne 0$ since otherwise
the identity is trivially satisfied. All the vectors $\gamma \in {\cal A}_{+}\cap \Pi$
can be splitted into equivalence classes $\Gamma_1,...,\Gamma_p$ according to the equivalence relation:
$\gamma \sim \gamma'$ if
\begin{equation}
\label{equiv}
\gamma' = \pm \gamma + \mu \alpha 
\end{equation}
for some $\mu.$ \footnote {Similar equivalence has been introduced by M.V.Feigin in the theory of trigonometric locus configurations
\cite{MF}.} The restrictions of $\cot(\gamma,x)$ and $\cot(\gamma',x)$
on the intersection line of $\Pi$ with the hyperplane $(\alpha,x)=0$ are the same up to a sign.
To take care of this sign let us introduce a skew-symmetric bilinear form $B = B_{a,b}(\alpha,\beta)$
related to a pair of vectors $a,b \in {\bf R}^n$ by the formula
$$B_{a,b}(\alpha,\beta) = (\alpha,a)(\beta,b) - (\alpha,b)(\beta,b).$$
It is easy to check that the identity (\ref{plane}) for the plane $\Pi$ is equivalent to
the set of relations
\begin{equation}
\label{B}
\sum\limits_{\gamma\in\Gamma_s, \gamma \ne \alpha} m_{\alpha}m_{\gamma} (\alpha, \gamma) B(\alpha, \gamma) = 0
\end{equation}
for all the equivalence classes $\Gamma_1,...,\Gamma_p.$
As a corollary we have the same relation on each plane $\Pi$:
\begin{equation}
\label{BP}
\sum\limits_{\beta \ne \alpha, \beta\in{\cal A}_{+}\cap \Pi} m_{\alpha}m_{\beta} (\alpha, \beta) B(\alpha, \beta) = 0
\end{equation}
and thus in the whole space:
\begin{equation}
\label{BS}
\sum\limits_{\beta \ne \alpha, \beta \in {\cal A}_{+}} m_{\alpha}m_{\beta} (\alpha, \beta) B(\alpha, \beta) = 0
\end{equation}
for any $\alpha \in {\cal A}_{+}.$

Let us introduce now the operators $M_{\alpha}$ in ${\bf R}^n$ by the relation
$$M_{\alpha}= m_{\alpha}\alpha\otimes\alpha,$$ 
where by definition $\alpha\otimes\beta (x) = (\beta,x) \alpha.$
Let $M$ be the sum 
$$M = \sum\limits_{\alpha \in {\cal A}_{+}}m_{\alpha}\alpha\otimes\alpha.$$

Now we need the following natural definition. We will call a configuration ${\cal A}$ {\it irreducible} if
it can not be splitted into two non-trivial orthogonal parts.

{\bf Lemma 2.} {\it The Main Identity (\ref{Main}) implies that all the operators
$M_{\alpha}, \alpha \in {\cal A}_{+}$ commute with $M$. 
For irreducible configurations this means that $M$ is a scalar operator:
$M = \mu I$ for some $\mu$.}

Indeed, the commutator $C_{\alpha}=[M, M_{\alpha}]$  is
$C_{\alpha} = [\sum\limits_{\beta \in {\cal A}_{+}}m_{\beta}\beta\otimes\beta, 
m_{\alpha}\alpha\otimes\alpha] =
- \sum\limits_{\beta \ne \alpha, \beta \in {\cal A}_{+}} m_{\alpha}m_{\beta} (\alpha, \beta) 
(\alpha\otimes\beta -\beta\otimes\alpha).$
It is easy to see that the relations (\ref{BS}) are equivalent to 
$(C_{\alpha} (a), b) =0$ for arbitrary $a,b \in {\bf R}^n$ which imply that
$C_{\alpha}=0.$ The commutativity of the operator $M$ with $M_{\alpha}$
means that any $\alpha \in {\cal A}$ is an eigenvector of the (self-adjoint) operator $M$.
If $M$ has two different eigenvalues than ${\cal A}$ can be splitted into two orthogonal parts
which is impossible in the irreducible case. This proves the Lemma.

Now everything is ready for the proof of our main result.
For a given configuration $\cal{A}$ let us define the set $\mathfrak{A}$
in the Euclidean space $V = {\bf R}^n  \simeq V^{*}$ by taking all the vectors of the 
form $\sqrt{m_{\alpha}} \alpha, \alpha \in \cal{A}_+.$

{\bf Lemma 3.} {\it If $\cal{A}$ is irreducible and satisfies the Main Identity 
then $\mathfrak{A}$ satisfies the relations (\ref{pi}) from the previous section.}

Indeed, due to lemma 2 in this case 
$G^{\mathfrak{A}}=\sum\limits_{\alpha \in {\cal A}_{+}}m_{\alpha}\alpha\otimes\alpha$
is proportional to the scalar product in ${\bf R}^n.$ The same is obviously true for $G_{\mathfrak{A}}$
so the relations (\ref{pi}) are reduced to (\ref{BP}). 

{\bf Remark.} Strictly speaking we should consider only positive multiplicities 
in order to stay on reals but one can easily check that the resulting function
$$F(x) =\sum\limits_{\alpha \in \cal {A}_{+}} m_{\alpha} (\alpha,x)^2 \, {\rm log} \, 
(\alpha,x)^2$$ is real even in the case of negative multiplicities and all the arguments
work in the complex case as well.

Combining these three lemmas with the Proposition 1 we have the proof
of the Theorem in the irreducible case. Reducible case then easily follows.

Now using the results of \cite{V1,V2} we have the following

{\bf Corollary.} {\it For any configuration $\cal{A}$ satisfying the Main Identity
the corresponding set $\mathfrak{A}$ is a $\vee$-system in the sense of \cite{V1}.
In particular the
following differential operators of the Knizhnik-Zamolodchikov type
\beq
\label{dkz}
\bigtriangledown_a = \partial_a - \sum\limits_{\alpha \in \cal{A}_{+}} 
m_{\alpha}\frac{(\alpha,a)}{(\alpha,x)}\alpha \otimes \alpha.
\eeq
commute and therefore define a flat connection in} ${\bf R}^n$ .

As the examples one can consider the following "locus configurations" $A_n(m)$ and $C_{n+1}(m,l)$ 
which have been introduced in \cite{VFCh, ChFV1,ChFV2}. They consist of the following vectors in ${\bf R}^{n+1}$:
$$
A_n(m)=
\left\{
\begin{array}{lll}
e_i - e_j, &  1\le i<j\le n, & {\rm with \,\, multiplicity \,\,}   m,\\
e_i - \sqrt{m}e_{n+1}, &  i=1,\ldots ,n  & {\rm with \,\, multiplicity \,\,}  1,
\end{array}
\right.
$$
and
$$
C_{n+1}(m,l) =
\left\{
\begin{array}{lll}
e_i\pm e_j, &  1\le i<j\le n, &  {\rm with \,\, multiplicity \,\,}   k,\\
2e_i, &  i=1,\ldots ,n  & {\rm with \,\, multiplicity \,\,}   m,\\
e_i\pm \sqrt{k}e_{n+1}, & i=1,\ldots ,n  &  {\rm with \,\, multiplicity \,\,}  1,\\
2\sqrt{k}e_{n+1} & {\rm with \,\, multiplicity \,\,}  l,\\
\end{array}
\right.
$$
where $k = \frac{2m+1}{2l+1}$.
The fact that they satisfy the Main Identity (for all values of the parameters) 
can be checked by direct calculation. Our Theorem thus explains the observation \cite{V1}
that the corresponding functions $F$ satisfy the generalised WDVV equation.

\section*{Concluding remarks.}
We have seen that the Main Identity is behind the relation between the locus configurations
and WDVV equation. Since there are locus configurations without this property this explains
also why this relation does not always work \cite{ChV}. An interesting problem is to classify
all locus configurations which satisfy the Main Identity. The fact that
this is true for the series $A_n(m)$ and $C_{n+1}(m,l)$ seems to
be very important. In particular, as it follows from our analysis in the previous section
these configurations (with integer parameters) are the "vector systems" in the Borcherds' sense \cite{B}.
Their role in Borcherds' theory deserves a special investigation.

We have already mentioned the relations with the theory of symmetric spaces
and the radial parts of the Laplace operators for the usual root systems.
It is interesting that the operators related to $A_n(m)$ and $C_{n+1}(m,l)$ also can be interpreted 
in a similar way using the Lie superalgebras and symmetric superspaces as it was recently
discovered by A.N. Sergeev \cite{Ser}. This relation gives some new interesting examples of the
integrable deformations of Calogero-Sutherland operators satisfying the Main Identity \cite{SV}
(and therefore new solutions for the WDVV equation).

\section*{ Acknowledgements.}
I am grateful to O.A. Chalykh, M.V. Feigin, V.V. Nikulin and A.N. Sergeev 
for stimulating discussions and useful comments.

This work has been partially supported by EPSRC.

\end{document}